# Evaluation of the impact of solder die attach versus epoxy die attach in a state of the art power package

Julia Czernohorsky[*], Bartosz Maj[*], Matthias Viering[*], Lance Wright[§], Gerry Balanon[#]

[*]Continental Teves AG, P.O. Box 90 01 20, 60441 Frankfurt, Germany
[§]Texas Instruments Inc., 12500 TI BLVD, MS 8761, Dallas, Texas 75243
[#]Texas Instruments Inc., 13560 North Central Expressway, MS3744, Dallas, Texas 75243

e-mail: julia.czernohorsky@contiautomotive.com, bartosz.maj@contiautomotive.com,
matthias.viering@contiautomotive.com, lwright@ti.com, g-balanon2@ti.com

*Abstract*—Subject of this paper is the thermal investigation of epoxy (EDA) and solder (SDA) die attaches by a comparison of an ASIC with multiple heat sources in different package assemblies. Static and transient thermal measurements and simulations were performed to investigate the thermal behavior of two samples in a state of the art QFP power package differing only in the die attach material (EDA and SDA).

I. INTRODUCTION

Modern day mixed signal devices are a combination of many different building blocks, of which some may contribute to the overall heat dissipation of the device, where as others are heat sensitive. Hence not only the absolute temperature but also the temperature distribution at junction level becomes a vital parameter for the electrical performance of such devices. A very important role in thermal considerations has to be attributed to the materials used on the component level and to the overall final module assembly. Due to this fact the intensified focus on thermal investigations is easily understandable. Additional factors such like increasing packing density, installation space limitations and cost factors add up to the importance of a good thermal design.

From the thermal performance point of view it is more and more important to estimate the impact of even smallest component changes for the whole module [1]. To fulfill this need such changes have to be evaluated also on the module level. Only such approach enables further optimization opportunities concerning the mechanical build up of the modules, where apart pure thermal considerations, also thermo-mechanical benchmarks have to be accounted for.

Subject of this paper is the thermal investigation of epoxy (EDA) and leadfree solder (SDA) die attaches by a comparison of the thermal performance of an ASIC with multiple heat sources in a state of the art QFP power package assembly. First of all the thermal behavior of the different die attach materials was investigated by thermal measurements using thermocouples and ESD diodes for temperature measurements on junction level. Afterwards thermal simulations were performed in order to evaluate the differences in temperature distribution of the two assemblies. After model verification based on the measurement results, steady state as well as transient simulations were performed and the heat propagation was investigated. The presented measurement and simulation results are discussed and finally some concluding remarks regarding the benefit of SDA devices are made.

II. MEASUREMENTS

*A. Measurement setup*

For the temperature investigations two samples in a QFP power package differing only in the die attach material (EDA and SDA) were populated on identical 6 layer application-like PCB's, which were equipped with a thermocouple. It was ensured that two assemblies of comparable die attach coverage were used. The temperature on junction level was determined with the help of ESD diodes. For both assemblies the same actuation was used to perform transient measurements.

For the temperature measurement different sensors are used. First of all both boards were equipped with a thermocouples to measure the temperature under the ASIC. Additionally some of the ESD diodes of the device were used to determine the temperature on junction level.

Before the use of the ESD diodes for temperature measurement, each diode has to be characterized in its temperature behavior. It is important during the characterization to evaluate the electrical offset of the diode voltage caused by powering on the device, thus biasing it. This evaluation has been done with a parameter analyzer for three temperatures (20°C, 60°C and 100°C). Important is to use a high time resolution assuming the electrical contribution is a very fast effect while the temperature rise is slower. Afterwards, the voltage offsets, depending on the used ESD structure, are subtracted from the measurement results to omit the electrical effects for the temperature rise measurement (see Fig.1).

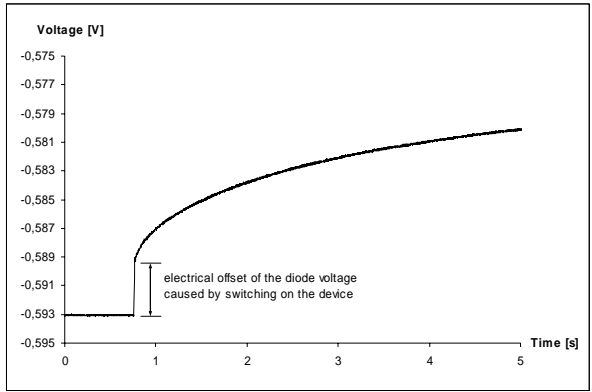

Fig.1: ESD diode characterization using the parameter analyzer. The measurement is done with 50μA load.

The ASIC's used for temperature investigations have over ten heat sources, of which some are permanently active and some can be dynamically switched on and off.

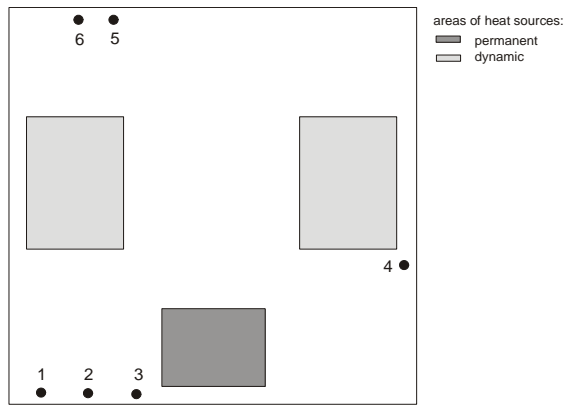

Fig.2: Schematical top view of the die. The dots indicate the positions of the six ESD diodes used to evaluate the temperature rise at the junction level of the ASIC.

The steady state case was investigated under two different load conditions. First minimal power dissipation within the permanent active heat sources was investigated, subsequently the power dissipation was increased to application case values. This was carried out to ensure an easier correlation between the measurement and simulation, neglecting as much as possible the effect of selfheating during the ESD diode characterization. As application specific a transient load condition with 10s activation of all dynamic sources (100% DC) was performed.

*B. Measurement results*

An overview of the accomplished measurements is given in this paragraph. Not all measurement data is presented, but only few chosen cases and temperature sensors.

The thermal conductivity of the solder die attach is about factor nine higher than the conductivity of the epoxy die attach, but due to the assembly process the thickness of the solder layer is about factor 2.5 greater. Thus the resulting thermal resistance of the solder die attach is about factor 3.5 smaller than the resistance of the epoxy die attach. Due to the difference in thermal resistances of the two package assemblies, a junction temperature of the SDA device of about 2°C and 10°C is expected for the two load conditions. But the steady state measurement results are showing only a difference of smaller 1°C between the EDA and SDA samples, while the power dissipation of the second load condition is about factor 7 higher.

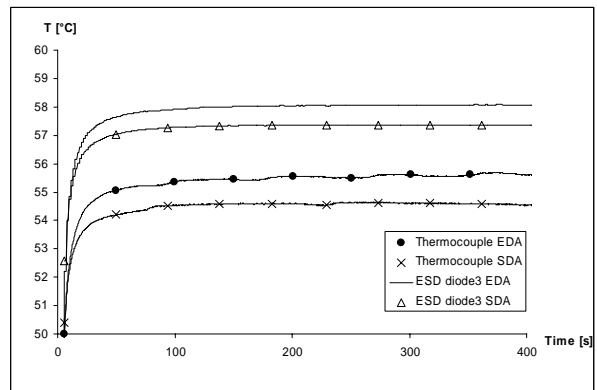

Fig.3: Steady state measurements with low power dissipation.

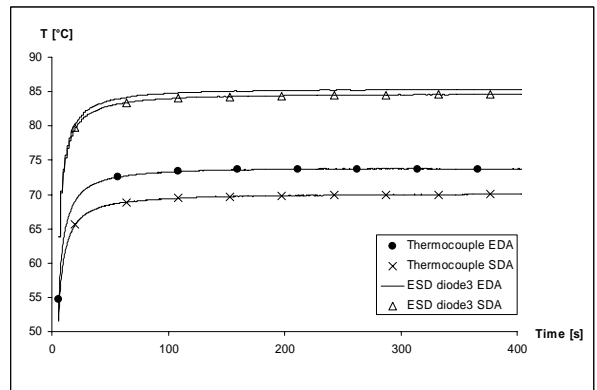

Fig.4: Steady state measurements with increased power dissipation.

The transient measurements preformed are showing additional impact of the assembly in the range of 3°C, and also a difference in the heat distribution in SDA devices, when comparing the results of ESD diodes 2 and 4 for both samples (Fig.6)

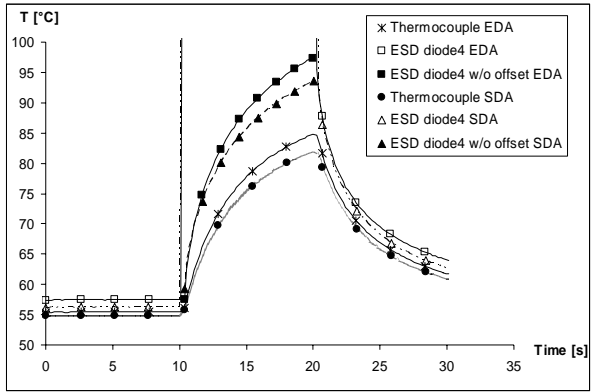

Fig.5: Transient response of the ESD diode 4 on the 10s load. The offset, caused when switching on the dynamic heat sources, is subtracted from the measurement results to omit the electrical effects.

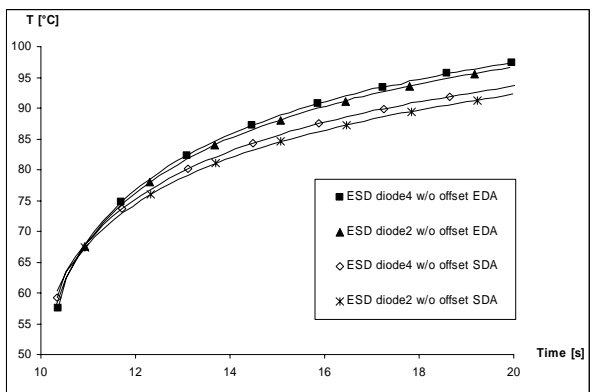

Fig.6: Transient response of the ESD diodes 2 and 4 (with offset correction) for both samples.

The fact that the measured temperature difference between EDA and SDA samples is much smaller than expected due to the difference in thermal resistances, is in good accordance with the already published information concerning die attach materials. Other investigations [2] showed that the impact of the variation of die attach materials is only significant in the range of lower thermal conductivities (< 7W/mK). Hence solder die attaches (with a large thermal conductivity) lead only to a relatively small decrease in junction temperature.

A smaller heat conducting volume in the package assembly with solder die attach would be an explanation for the small temperature difference between EDA and SDA samples. This assumption was investigated with the help of thermal simulations.

## III. SIMULATION

### A. Model verification by measurements

The FEA is performed with a commercial analysis software (ANSYS 11.0). The FEA model used for thermal simulations is representing the measurement setup. The ASIC is placed in the center of the printed circuit board. As electrical connections of the package are of negligible influence, pins and bond wires are not represented in the model. Simplifications are also used for the 6 layer PCB. Homogeneous PCB material is assumed by calculating average material properties based on layer setup and copper proportions of the single layers. The ASIC is connected to the PCB via soldering. Underneath the ASIC the PCB has a bore which is solder filled (Fig.7).

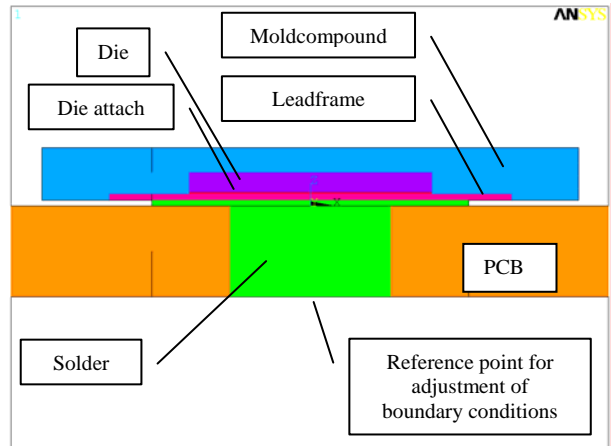

Fig.7: Section through package and PCB

Uniform and constant convection is applied to top and bottom side of the PCB as well as the outside of the IC package. The convectional film coefficient is adjusted to steady state measurement results of the thermocouple at PCB bottom side (Fig.7).

Steady state thermal load is applied to the permanent heat sources, transient loads occur at the dynamic heat sources. Loads are applied as heatfluxes to the surface of the die.

### B. Simulation results

The assumption that a narrower temperature distribution of SDA devices leads to smaller temperature differences than expected, is further investigated with the help of steady state and transient simulation results.

In order to investigate the difference in the heat conducting volume the device is cut along three axes (Fig.8): horizontal (1) and vertical (2) through the dynamic heat sources and also vertical through the permanent heat source with the highest power dissipation (3). Along those axes the temperature distribution is investigated at different levels of the assembly: e.g. junction level, middle of the die and top of the die attach.

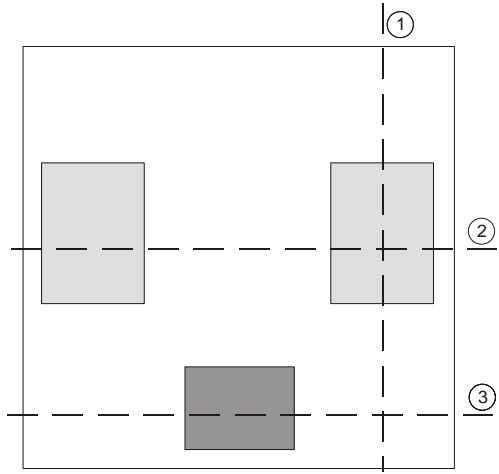

Fig.8: Temperature distribution is investigated along three axes.

The curves of Fig.9 show the temperature distribution along axis 3 for the steady state simulation results.

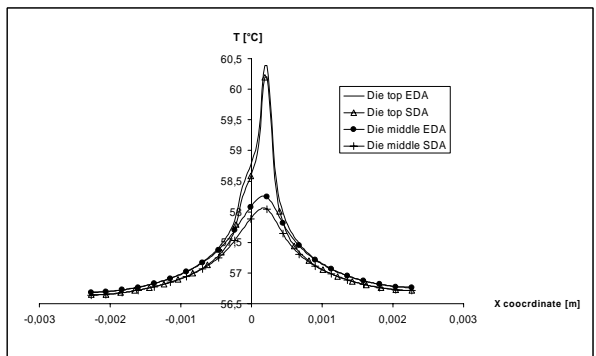

Fig.9: Temperature distribution along axis 3.

Considering the distance of equipotential lines on junction level – here from 100% to 95% of $T_{max}$ – a difference in the spread in x-direction beyond the area of the heat source can be observed.

Due to the higher power dissipation this effect becomes more apparent considering the transient simulation results. The temperature distribution along axis 1 and 2 after 10s activation of the dynamic heat sources is shown in Fig.10 and 11.

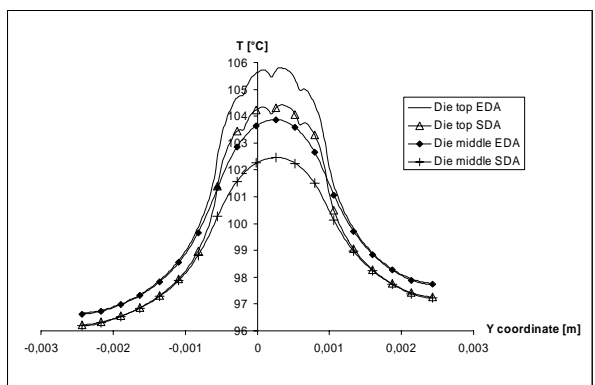

Fig.10: Temperature distribution along axis 1.

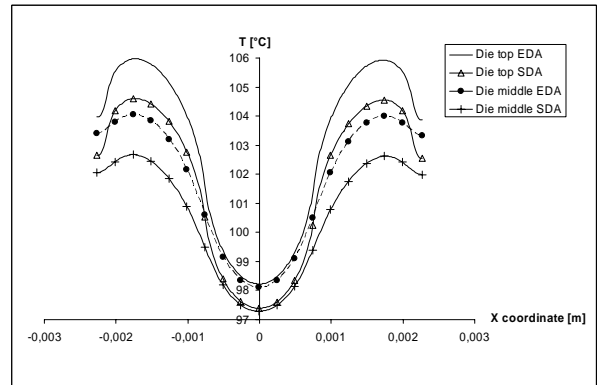

Fig.11: Temperature distribution along axis 2.

Considering the distance in y-direction at $T_{max}$ and at 95% of $T_{max}$ a difference of about 40µm between EDA and SDA assemblies can be detected (Fig12). The decline in temperature (in % of $T_{max}$) with increasing distance from the heat source is faster at the SDA device, thus the area of high temperatures around the heat source is smaller. The difference in spread in y-direction is most apparent in the middle of the die. At the top of the die attach this effect seems to interfere with the heat accumulation caused by the abrupt rise in thermal resistance when reaching the die attach.

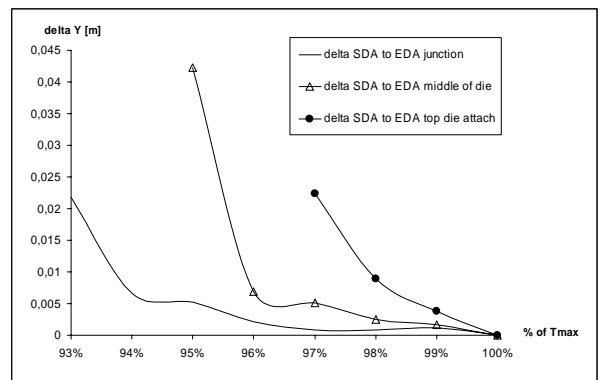

Fig.12: Temperature spread in y-direction on junction level, in the middle of the die and at the top of the die attach.

The setup used for these investigations was not beneficial in order to investigate the differences in the heat conducting volume of EDA and SDA devices. The observed differences will probably become more apparent if a heatsink, capable to dissipate the power from the ASIC, is directly connected to the leadframe. Further investigations using such a heatsink concept are expedient for a more detailed research of the effects which are presented in this paper. [3]

V. CONCLUSIONS

This paper demonstrates that the die attach material does not have a significant impact on the absolute maximum temperatures, but the solder die attach causes a narrower temperature distribution on and across the die. Thus a SDA assembly provides the possibility to reduce the spacing between heat sources and heat-sensitive components of the device. The overall benefit of this is highly dependent on the boundary conditions on the interface between assembly and ambient, meaning the system heatsink design. As already indicated the setup used for this assessment was not beneficial to evaluate the maximum possible thermal advantage of the SDA vs. EDA assembly. This assesment will be done with a application near assembly where the ASIC leadframe is directly connected to a system heatsink and not soldered on a heatsinkless PCB. Additionally the SDA assembly has to be benchmarked also under thermo-mechanical considerations regarding environmental influences (e.g. thermal cycles, power cycles, leadframe warpage, corrosion, Moisture Sensivity Level testing and autoclave).